\documentclass[12pt]{article}
\usepackage{epsf}

\begin{document}

\begin{titlepage}
\begin{flushright}
CLNS~02/1794\\
{\tt hep-ph/0207327}\\[0.2cm]
\today
\end{flushright}

\vspace{1.0cm}
\begin{center}
\Large\bf Two-body Modes of B Mesons and\\
The CP-b Triangle\footnote{Invited plenary talk presented at the 
$6^{th}$ International Workshop on {\em Heavy Quarks and Leptons}, 
Vietri sul mare, Salerno, Italy (May 27--June 1, 2002), and at the 
International High-Energy Physics Conference on {\em Quantum 
Chromodynamics}, Montpellier, France (July 2--9, 2002)}
\end{center}

\vspace{1.0cm}
\begin{center}
Matthias Neubert\\[0.1cm]
{\sl F.R. Newman Laboratory for Elementary-Particle Physics\\
Cornell University, Ithaca, NY 14853, USA}
\end{center}

\vspace{1.0cm}
\begin{abstract}
\vspace{0.2cm}\noindent 
The study of charmless two-body decays of $B$ mesons is currently one 
of the hottest topics in $B$ physics. QCD factorization provides the 
theoretical framework for a systematic analysis of such decays. A 
global fit to $B\to\pi K,\pi\pi$ branching fractions, combined with 
knowledge on $|V_{ub}|$, establishes the existence of a CP-violating 
phase in the bottom sector of the CKM matrix and tends to favor values 
of $\gamma$ near $90^\circ$, somewhat larger than those suggested by 
the standard analysis of the unitarity triangle. A novel construction 
of the unitarity triangle is presented, which is independent of 
$B$--$\bar B$ and $K$--$\bar K$ mixing. It can provide stringent tests 
of the Standard Model with small theoretical uncertainties.
\end{abstract}
\vfill
\end{titlepage}

\section{Introduction}

\begin{figure}[h]
\epsfxsize=12cm
\centerline{\epsffile{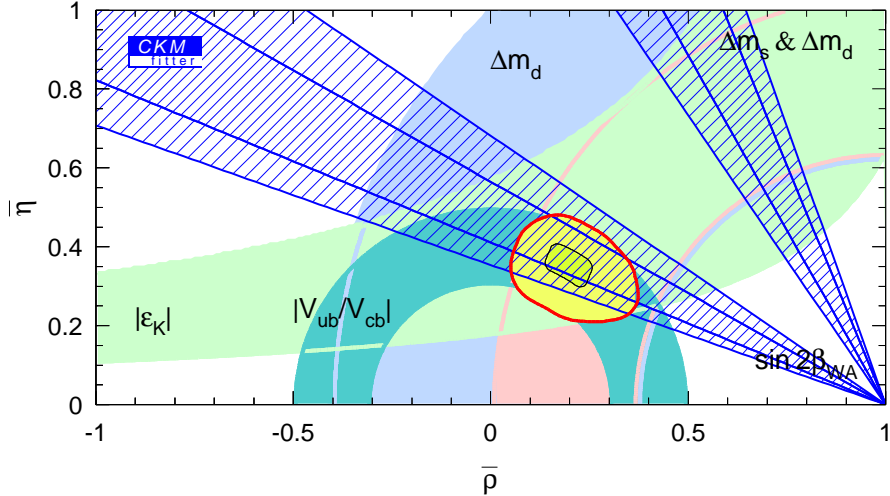}}
\centerline{\parbox{15cm}{\caption{\label{fig:UTfit}
Standard constraints on the apex $(\bar\rho,\bar\eta)$ of the unitarity 
triangle \protect\cite{Hocker:2001xe}.}}}
\end{figure}

Measurements of $|V_{ub}|$ in semileptonic decays, $|V_{td}|$ in 
$B$--$\bar B$ mixing, and $\mbox{Im}(V_{td}^2)$ from CP violation in 
$K$--$\bar K$ and $B$--$\bar B$ mixing have firmly established the 
existence of a CP-violating phase in the CKM matrix. The present 
situation, often referred to as the ``standard analysis'' of the 
unitarity triangle, is summarized in Figure~\ref{fig:UTfit}. 

Three comments are in order concerning this analysis:
\begin{enumerate}
\item
The measurements of CP asymmetries in kaon physics ($\epsilon_K$ and 
$\epsilon'/\epsilon$) and $B$--$\bar B$ mixing ($\sin2\beta$) probe the 
imaginary part of $V_{td}$ and so establish CP violation in the top 
sector of the CKM matrix.\footnote{Here I adopt the standard phase 
conventions for the CKM matrix. The corresponding convention-independent 
statement is that $\mbox{Im}[(V_{td}V_{ts}^*)/(V_{cd}V_{cs}^*)]\ne 0$ 
and $\mbox{Im}[(V_{td}V_{tb}^*)/(V_{cd}V_{cb}^*)]\ne 0$.}
The CKM model predicts that the imaginary part of $V_{td}$ is related, by 
three-generation unitarity, to the imaginary part of $V_{ub}$, and that 
those two elements are (to an excellent approximation) the only sources 
of CP violation in flavor-changing processes. In order to test this 
prediction, the next step must be to explore the CP-violating phase 
$\gamma=\mbox{arg}(V_{ub}^*)$ in the bottom sector of the CKM matrix. 
In this talk I argue that the analysis of charmless hadronic $B$ decays
has by now established unambiguously that $\mbox{arg}(V_{ub}^*)\ne 0$.
\item
With the exception of the $\sin2\beta$ measurement, the standard 
analysis is limited by large theoretical uncertainties, which dominate 
the widths of the various bands in the figure. These uncertainties enter 
via the calculation of hadronic matrix elements. I will discuss some 
novel methods to constrain the unitarity triangle using charmless 
hadronic $B$ decays, which are afflicted by smaller hadronic 
uncertainties and hence provide powerful new tests of the Standard Model,
which can complement the standard analysis.
\item
With the exception of the measurement of $|V_{ub}|$ in semileptonic $B$ 
decays, the standard constraints are sensitive to meson--antimeson 
mixing. Mixing amplitudes are of second order in weak interactions and 
hence might be most susceptible to effects from physics beyond the 
Standard Model. The new constraints on $(\bar\rho,\bar\eta)$ discussed 
below allow a construction of the unitarity triangle that is 
over-constrained and independent of $B$--$\bar B$ and $K$--$\bar K$ 
mixing. It is in this sense orthogonal to the standard analysis.
\end{enumerate}

\begin{figure}
\epsfxsize=12cm
\centerline{\epsffile{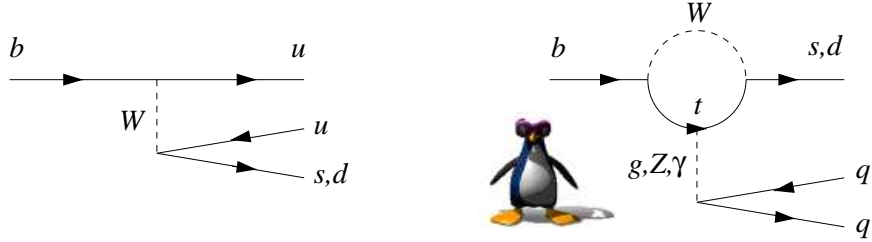}}
\centerline{\parbox{15cm}{\caption{\label{fig:topol}
Tree and penguin topologies in charmless hadronic $B$ decays.}}}
\end{figure}

The phase $\gamma$ can be probed via tree--penguin interference in 
decays such as $B\to\pi K,\pi\pi$, for which the underlying flavor 
topologies are illustrated in Figure~\ref{fig:topol}. Experiment teaches 
us that amplitude interference is sizable in these decays. Information 
about $\gamma$ can be obtained not only from the measurement of direct 
CP asymmetries ($\sim\sin\gamma$), but also from the study of CP-averaged 
branching fractions ($\sim\cos\gamma$). The challenge is, of course, to
gain theoretical control over the hadronic physics entering the 
tree-to-penguin ratios in the various decays.

\section{QCD Factorization}

Hadronic weak decays simplify greatly in the heavy-quark limit 
$m_b\gg\Lambda_{\rm QCD}$. The underlying physics is that a fast-moving 
light meson produced by a point-like source (the effective weak 
Hamiltonian) decouples from soft QCD interactions 
\cite{Bjorken:1989kk,Dugan:1991de,Politzer:1991au}. A systematic 
implementation of this color transparency argument is provided by the 
QCD factorization approach \cite{Beneke:1999br,Beneke:2001ev}. This 
scheme makes rigorous predictions in the heavy-quark limit, some of which 
have been proven to all orders of perturbation theory \cite{Bauer:2001cu}. 
One can hardly overemphasize the importance of controlling nonleptonic 
decay amplitudes in the heavy-quark limit. While a few years ago reliable
calculations of such amplitudes appeared to be out of reach, we are now 
in a situation where hadronic uncertainties enter only at the level of 
power corrections suppressed by the heavy $b$-quark mass. 

\begin{figure}
\centerline{\epsfxsize=6.6cm\epsffile{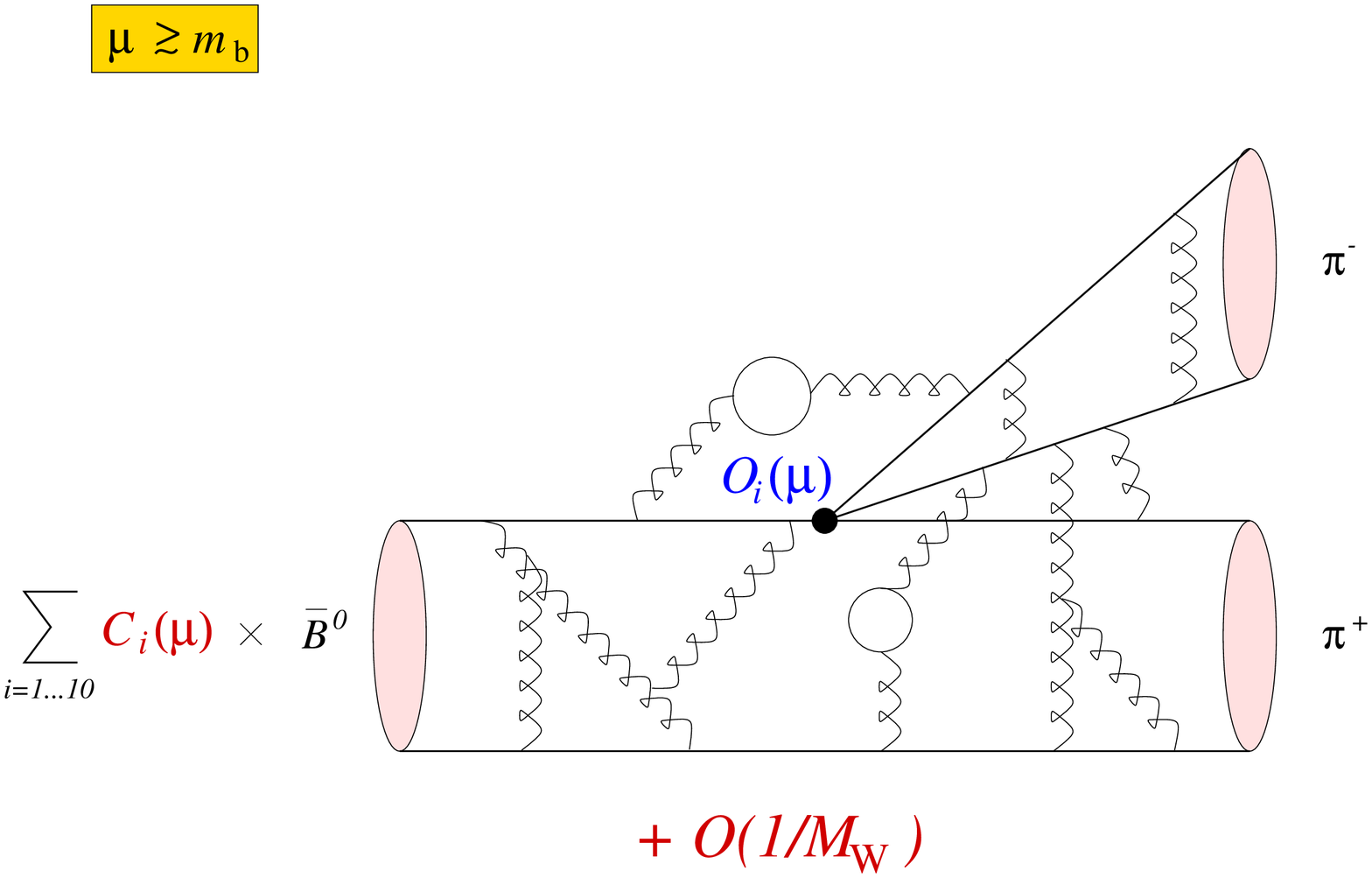}\quad%
\epsfxsize=8.7cm\epsffile{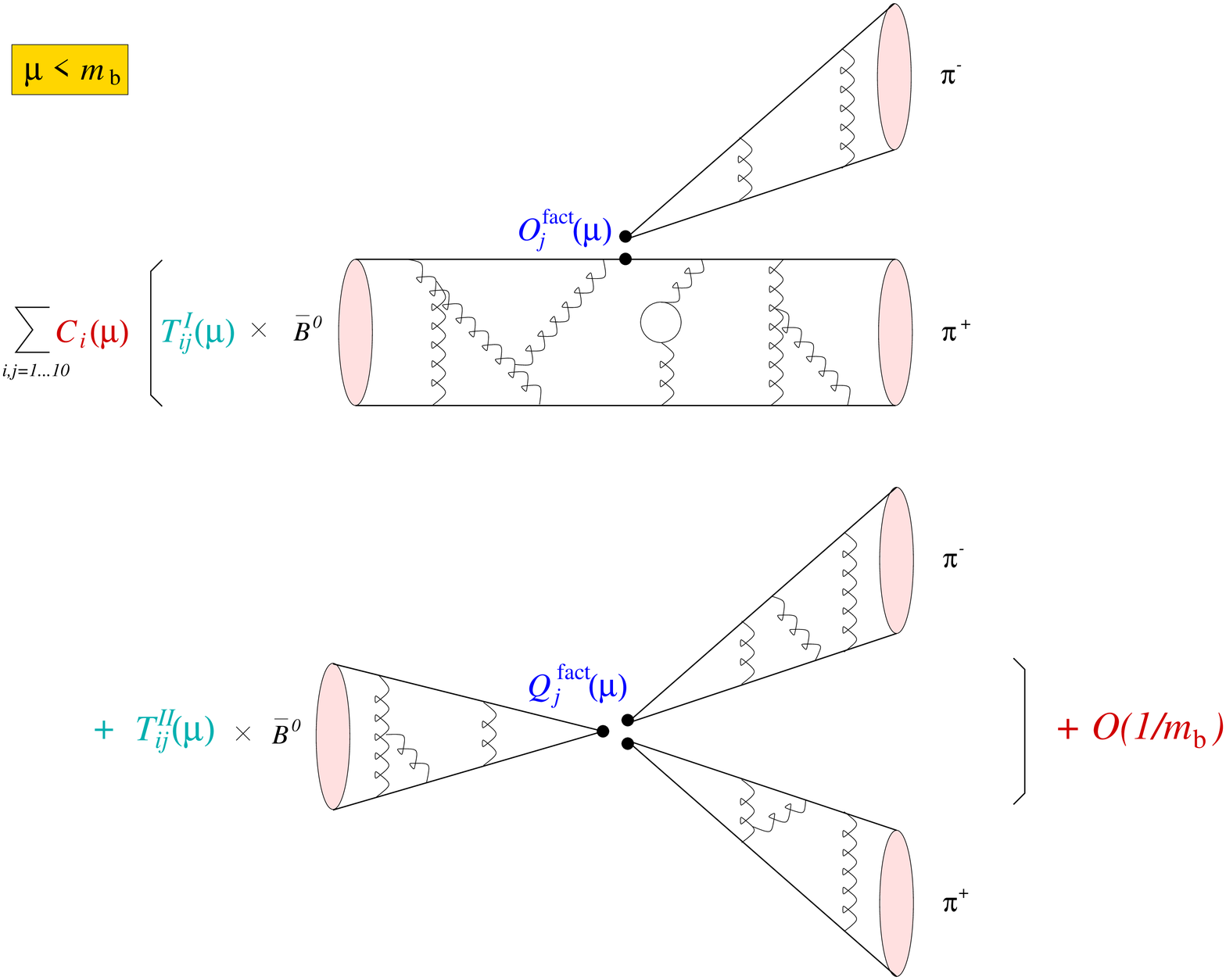}}
\centerline{\parbox{15cm}{\caption{\label{fig:fact}
Factorization of short- and long-distance contributions into running
couplings and hadronic matrix elements. Left: Integrating out hard 
gluons ($k\sim M_W$) in the construction of the effective weak 
Hamiltonian. Right: Integrating out hard gluons ($k\sim m_b$) in the 
construction of the effective, factorized transition operator in QCD
factorization.}}}
\end{figure}

The workings of QCD factorization are illustrated in 
Figure~\ref{fig:fact}. The graph on the left shows how in the familiar 
construction of the effective weak Hamiltonian hard gluon effects with 
virtualities $\mu\gg m_b$ can be calculated and factorized into Wilson 
coefficients $C_i(\mu)$. The graphs on the right illustrate how, in a 
similar way, hard gluon effects with $\mu\sim m_b$ can be calculated 
and factorized into perturbative hard-scattering kernels $T_{ij}(\mu)$. 
What remains after this step are factorized decay amplitudes, in which 
all gluon exchange between the emission meson at the ``upper vertex'' 
and the remaining hadronic system are integrated out. In the 
heavy-quark limit, such ``nonfactorizable gluons'' are hard because of 
color transparency. Note that this does not imply that nonleptonic 
amplitudes in the heavy-quark limit are perturbative. (In this respect, 
our approach is more general that the pQCD scheme \cite{Keum:2001ph}.)
Important nonperturbative effects remain, which can be parameterized in
terms of meson decay constants, $B\to M$ transition form factors, and
meson light-cone distribution amplitudes. These quantities are an input
to the factorization formula, ideally taken from experiment. Theoretical 
expressions for decay amplitudes obtained using the QCD factorization 
approach are complicated and depend on many input parameters. When 
discussing the theoretical uncertainties and limitations of this scheme 
it is important to distinguish between different classes of parameters. 
In order of phenomenological importance, these are Standard Model 
parameters ($\bar\rho$, $\bar\eta$, $m_s$, $m_c$, $\alpha_s$), the 
renormalization scale ($\mu$), hadronic quantities that can (at least in 
principle) be determined from data (decay constants, transition form 
factors), and hadronic quantities that can only indirectly be 
constrained by data (light-cone distribution amplitudes).

The most important question with regard to phenomenological applications
of QCD factorization is that about the numerical size of power 
corrections. While the importance of the heavy-quark limit to the 
workings of factorization is evident from a comparison of 
nonfactorizable effects seen in kaon, charm and beauty decays 
\cite{Neubert:2001sj}, and while there is a lot of evidence (from 
spectroscopy, exclusive semileptonic decays, and various inclusive 
decays) that power corrections are small at the $b$-quark scale, it is 
nevertheless important to address the issue of power corrections in a 
systematic way. Much effort has been devoted in the past few years to 
the study of power-suppressed effects, which in general violate 
factorization. The most important power corrections are proportional to 
the ratios $2 m_K^2/(m_s m_b)$ or $2m_\pi^2/(m_q m_b)$ with $q=u,d$, which 
are inversely proportional to light-quark masses. Such twist-3 corrections 
make up for a significant portion of the penguin amplitudes in $B$ decays 
into light pseudoscalar mesons. It is important that these penguin 
contributions are calculable despite their power suppression and hence can 
be included reliably \cite{Beneke:2001ev}. At the same order, there appear 
logarithmically divergent twist-3 corrections to the leading-twist hard 
spectator interactions. These corrections are universal, and their effect 
can be absorbed into a redefinition of a single hadronic parameter 
$\lambda_B$.

Perhaps the largest uncertainty from power corrections is due to weak 
annihilation contributions, for which both of the valence quarks of the 
initial $B$ meson participate in the weak interactions
\cite{Keum:2001ph,Cheng:2000hv}. Annihilation amplitudes violate
factorization and thus cannot be reliably computed using the QCD
factorization approach. Although we find that with default parameter 
values the annihilation amplitudes are typically small, their effects 
can become sizable when the large model uncertainties in their 
estimate are taken into account \cite{Beneke:2001ev}. Other types of 
power corrections, such as soft nonfactorizable gluon exchange, have 
been investigated using QCD sum rules \cite{Khodjamirian:2001mi} and 
the renormalon calculus \cite{Burrell:2001pf,Becher:2001hu}. No large 
corrections of this type have been identified.

While it is a conceptual challenge to gain a better control over the
leading power corrections to QCD factorization, perhaps using the 
framework of the soft-collinear effective theory 
\cite{Bauer:2000yr,Chay:2002vy,Beneke:2002ph}, it is important that 
this approach makes many testable predictions. Their comparison with 
experimental data can teach us a lot about the importance of 
power-suppressed effects.

\section{Testing Factorization in \boldmath$B\to\pi K,\pi\pi$ 
Decays\unboldmath}
\label{sec:tests}

Deriving constraints on the unitarity triangle from charmless hadronic
$B$ decays requires controlling the interference of tree and penguin
topologies. This means that one must be able to predict not only the 
magnitudes of these contributions, but also their relative 
strong-interaction phase. Fortunately, the crucial aspects of such 
calculations can be tested using experimental data.

The magnitude of the leading $B\to\pi\pi$ tree amplitude can be probed in
the decays $B^\pm\to\pi^\pm\pi^0$, which to an excellent approximation 
do not receive any penguin contributions. The QCD factorization approach 
makes an absolute prediction for the corresponding branching ratio
\cite{Beneke:2001ev},
\[
   \mbox{Br}(B^\pm\to\pi^\pm\pi^0)
   = \Big[ 5.3_{\,-0.4}^{\,+0.8}\,(\mbox{pars.})
   \pm 0.3\,(\mbox{power}) \Big] \cdot 10^{-6}\times 
   \left[ \frac{|V_{ub}|}{0.0035}\,\frac{F_0^{B\to\pi}(0)}{0.28}
   \right]^2 ,
\]
which compares well with the experimental result 
$(4.9\pm 1.1)\times 10^{-6}$ \cite{newdata}. The theoretical 
uncertainties quoted are due to input parameter variations and to the 
modeling of power corrections. An additional uncertainty comes from the 
present error on $|V_{ub}|$ and the $B\to\pi$ form factor.

The magnitude of the leading $B\to\pi K$ penguin amplitude can be probed 
in the decays $B^\pm\to\pi^\pm K^0$, which to an excellent approximation 
do not receive any tree contributions. Combining it with the measurement
of the tree amplitude just described, a tree-to-penguin ratio can be 
determined via the relation
\[
   \varepsilon_{\rm exp} = \left| \frac{T}{P} \right|
   = \tan\theta_C\,\frac{f_K}{f_\pi} \left[ 
   \frac{2\mbox{Br}(B^\pm\to\pi^\pm\pi^0)}
        {\mbox{Br}(B^\pm\to\pi^\pm K^0)} \right]^{1/2}\!
   = 0.205\pm 0.025 \,.
\]
The experimental value of this ratio is in good agreement with the 
theoretical prediction $\varepsilon_{\rm th}=0.23\pm 0.04\,(\mbox{pars.})
\pm 0.04\,(\mbox{power})\pm 0.05\,(V_{ub})$ \cite{Beneke:2001ev}, which 
is independent of form factors but proportional to $|V_{ub}/V_{cb}|$. 
This is a highly nontrivial test of the QCD factorization approach. 
Recall that when the first measurements of charmless hadronic decays 
appeared several authors remarked that the penguin amplitudes were much 
larger than expected based on naive factorization models. We now see 
that QCD factorization naturally reproduces the correct magnitude of the 
tree-to-penguin ratio. This observation also shows that there is no need 
to supplement the QCD factorization predictions in an ad hoc way by 
adding enhanced phenomenological penguin amplitudes, such as the 
``nonperturbative charming penguins'' introduced in 
\cite{Ciuchini:1997hb}. 

QCD factorization predicts that most strong-interaction phases in 
charmless hadronic $B$ decays are parametrically suppressed in the 
heavy-quark limit, because
\[
   \sin\phi_{\rm st} = O[\alpha_s(m_b),\Lambda_{\rm QCD}/m_b] \,.
\]
This implies small direct CP asymmetries since, e.g., 
$A_{\rm CP}(\pi^+ K^-)\!\approx\!-2\,|T/P|\sin\gamma\,\sin\phi_{\rm st}$. 
The suppression results as a consequence of systematic cancellations of 
soft contributions, which are missed in phenomenological models of 
final-state interactions. In other schemes the strong-interaction phases 
are predicted to be larger, and therefore larger CP asymmetries are 
expected. Present data show no evidence for large direct CP asymmetries 
in charmless decays \cite{newdata}, but the errors are still too large 
to distinguish between different theoretical predictions. An important
exception is the direct CP asymmetry for the decays $B\to\pi^\pm K^\mp$, 
which is already measured with high precision. The current world average, 
$A_{\rm CP}(\pi^+ K^-)=-0.05\pm 0.05$ \cite{newdata}, implies a rather 
small value of the corresponding strong-interaction phase, which is 
consistent with the expectation that this phase be suppressed in the 
heavy-quark limit. Specifically, for $\gamma$ in the range between 
$60^\circ$ and $90^\circ$, I obtain $\phi_{\rm st}=(8\pm 10)^\circ$. 
Simple physical arguments suggest that the relevant strong-interaction 
phases in the decays $B\to\pi^\pm K^\mp$ and $B^\mp\to\pi^0 K^\mp$ 
should be very similar \cite{Gronau:1998ep}. This observation will 
become important below.

\section{Establishing CP Violation in the Bottom Sector}

\begin{figure}[t]
\centerline{\epsfxsize=7.5cm\epsffile{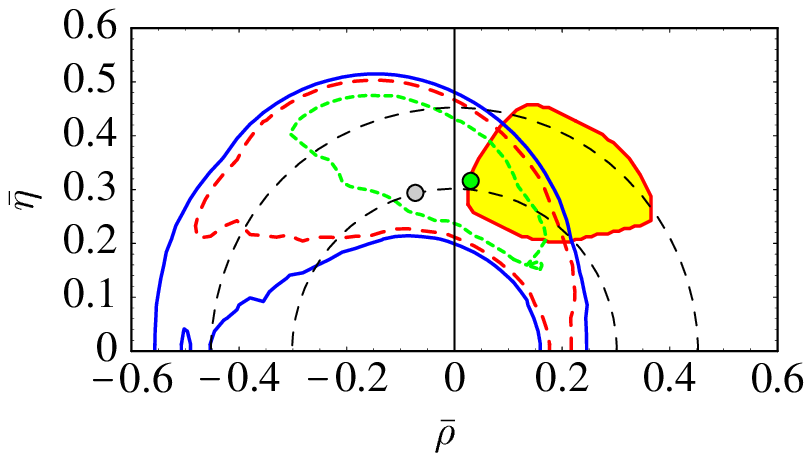}\quad%
\epsfxsize=7.5cm\epsffile{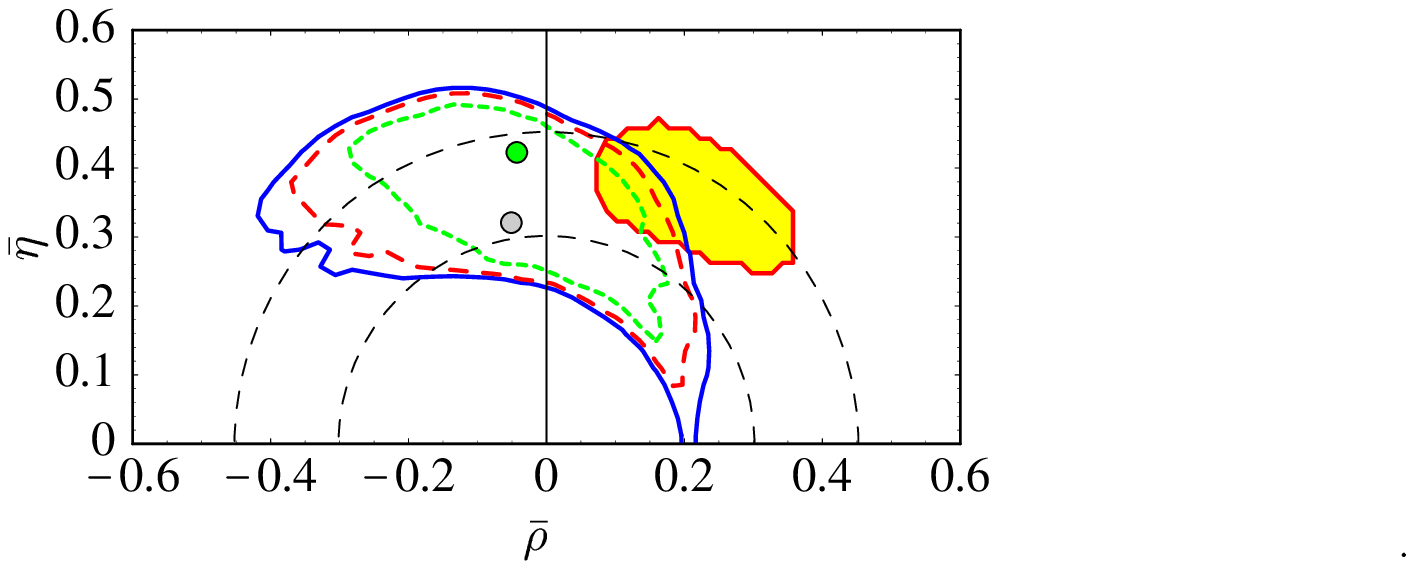}}
\centerline{\parbox{15cm}{\caption{\label{fig:rhoeta}
95\% (solid blue), 90\% (dashed red) and 68\% (short-dashed green) 
confidence level contours in the $(\bar\rho,\bar\eta)$ plane obtained 
from rare hadronic $B$ decays (dark green dot = overall best fit; light 
gray dot = best fit for the default parameter set). The circled yellow 
region shows the result of the standard CKM fit. Left: status in spring 
2001; right: update for summer 2002.}}}
\end{figure}

Various ratios of CP-averaged $B\to\pi K,\pi\pi$ branching fractions 
exhibit a strong dependence on $\gamma$ and $|V_{ub}|$. It is thus 
possible to derive constraints on $\bar\rho$ and $\bar\eta$ from a 
global analysis of the data in the context of the QCD factorization 
approach, provided conservative error estimates for power corrections 
are included. A comprehensive discussion of such an analysis was 
presented in \cite{Beneke:2001ev}, to which I refer the reader for 
details. The original result obtained in that paper is reproduced in 
the left plot in Figure~\ref{fig:rhoeta}. It reflects the status of 
the data as of spring 2001. The right plot shows an update of this 
analysis using the latest experimental data \cite{newdata}. I have 
also updated two input parameters in order to take into account recent 
theoretical developments. The new analysis uses $m_s=(100\pm 25)$\,MeV 
at $\mu=2$\,GeV, and $f_B=(200\pm 30)$\,MeV. The values adopted in 
\cite{Beneke:2001ev} were $m_s=(110\pm 25)$\,MeV and 
$f_B=(180\pm 40)$\,MeV.

The fit is excellent, which $\chi^2=0.5$ for three degrees of freedom. 
There is no problem in accounting for all of the experimental data 
simultaneously. The inclusion of model estimates of weak annihilation 
effects enlarges the allowed regions in the 
$(\bar\rho,\bar\eta)$ plane but is not required to fit the data.
Leaving out all annihilation contributions, one still obtains a good 
fit ($\chi^2=0.7$) and similar best-fit values for the Wolfenstein 
parameters. The comparison of the two plots shown in the figure 
indicates the effect of the increase in experimental precision between 
spring 2001 and summer 2002. The most important conclusion from this 
analysis is that, with the new data, the combination of results from 
rare hadronic $B$ decays with the $|V_{ub}|$ measurement in 
semileptonic decays (dashed circles) excludes $\bar\eta=0$ and so 
establishes the existence of a CP-violating phase in the bottom sector 
of the CKM matrix. 

The allowed regions obtained from the fit to charmless hadronic decays 
are compatible with the standard fit (shown by the yellow region), but 
tend to favor larger $\gamma$ values. This tendency has been reinforced 
with the new data. The same trend is seen in an analysis that does not 
rely on QCD factorization but instead employs general amplitude 
parameterizations and flavor symmetries \cite{Fleischer:2002zv}. It is 
tantalizing to speculate about the possible origin of a (still 
hypothetical) disagreement between the allowed $(\bar\rho,\bar\eta)$ 
regions obtained from the standard analysis and from charmless hadronic 
$B$ decays. A conventional explanation of such a discrepancy might be 
that the errors in lattice calculations of the relevant matrix elements 
for $B_d$--$\bar B_d$ and $B_s$--$\bar B_s$ mixing have been 
underestimated. In fact, in a recent paper the value 
$\xi=(f_{B_s}\sqrt{B_s})/(f_{B_d}\sqrt{B_d})=1.32\pm 0.10$ was obtained 
\cite{Kronfeld:2002ab}, which is significantly larger than the result 
$\xi=1.15\pm 0.05$ used in previous analyses of the unitarity triangle. 
With such large $\xi$, values of $\gamma$ in the vicinity of $90^\circ$ 
are no longer excluded by the $\Delta m_s/\Delta m_d$ bound. 

A more exciting possibility is, of course, to invoke New Physics to 
explain the discrepancy. Assume first that in charmless hadronic $B$ 
decays one probes the true value of the CKM phase $\gamma$. In this 
case a discrepancy with the standard analysis would most likely be due 
to a New Physics contribution to $B$--$\bar B$ mixing. For instance, 
there could be New Physics affecting $B_s$--$\bar B_s$ mixing. 
Eliminating the corresponding constraint from the standard analysis one 
finds that larger values of $\gamma$ are allowed. This possibility will 
hopefully soon be checked, when $B_s$--$\bar B_s$ mixing will be 
explored at the Tevatron. Alternatively, there could be New Physics 
affecting $B_d$--$\bar B_d$ mixing. In this case one should eliminate 
the constraints arising from the measurements of $\Delta m_d$, 
$\Delta m_s/\Delta m_d$, and $\sin2\beta$ from the standard analysis. 
Then only the constraints from $K$--$\bar K$ mixing and semileptonic 
$B$ decays remain, which allow for large values of $\gamma$. A different 
possibility would be that the mixing amplitudes are unaffected by New 
Physics, but that there exist non-standard contributions to $b\to s$ 
or $b\to d$ FCNC transitions, e.g.\ from penguin and box diagrams 
involving the exchange of new heavy particles. (A more exotic model with 
light SUSY particles has also been considered \cite{Becher:2002ue}.) In 
this case, $\gamma$ measured in $B\to\pi K,\pi\pi$ decays would be a 
combination of the CKM angle and some new CP-violating phase. Many 
examples of New Physics models that could yield a significant additional 
phase have been explored in \cite{Grossman:1999av}. A clean test of this 
possibility would be the measurement of the time-dependent CP asymmetry 
in $B\to\phi K_S$ decays, which in the Standard Model is due to the 
interference of a (real) $b\to s\bar s s$ penguin amplitude with the 
$B_d$--$\bar B_d$ mixing amplitude. If there was a New Physics phase 
$\phi_{\rm NP}$ of the penguin amplitude, then the CP asymmetry in 
$B\to\phi K_S$ would measure $\sin2(\beta+\phi_{\rm NP})$, which when 
compared with the value of $\sin2\beta$ measured in $B\to J/\psi\,K_S$ 
decays would reveal the existence of the phase $\phi_{\rm NP}$
\cite{Grossman:1996ke}. Note that this strategy would not be invalidated 
even if there was a New Physics contribution to $B_d$--$\bar B_d$ 
mixing. In this case $\beta$ would no longer be given by the CKM phase, 
but this effect would cancel out in the comparison of the two decay 
modes.

\section{A Mixing-Independent Construction of The Unitarity Triangle}

If the trend toward larger $\gamma$ values revealed by the analysis of
charmless hadronic $B$ decays persists, one will want to check the 
compatibility with the standard analysis using measurements whose 
theoretical interpretation is ``clean'' in the sense that it only 
relies on assumptions that can be tested using experimental data. Here 
I propose a novel construction of the unitarity triangle (which I call 
the CP-$b$ triangle, because it establishes a CP-violating phase in the 
$b$ sector of the CKM matrix) which has this property, is over-determined, 
and can be performed using already existing data. Most importantly, this 
construction is insensitive to potential New Physics effects in 
$B$--$\bar B$ or $K$--$\bar K$ mixing. I will argue that the theoretical 
uncertainties limiting this construction are considerably smaller than 
for the standard analysis.

\begin{figure}
\centerline{\epsfxsize=5cm\epsffile{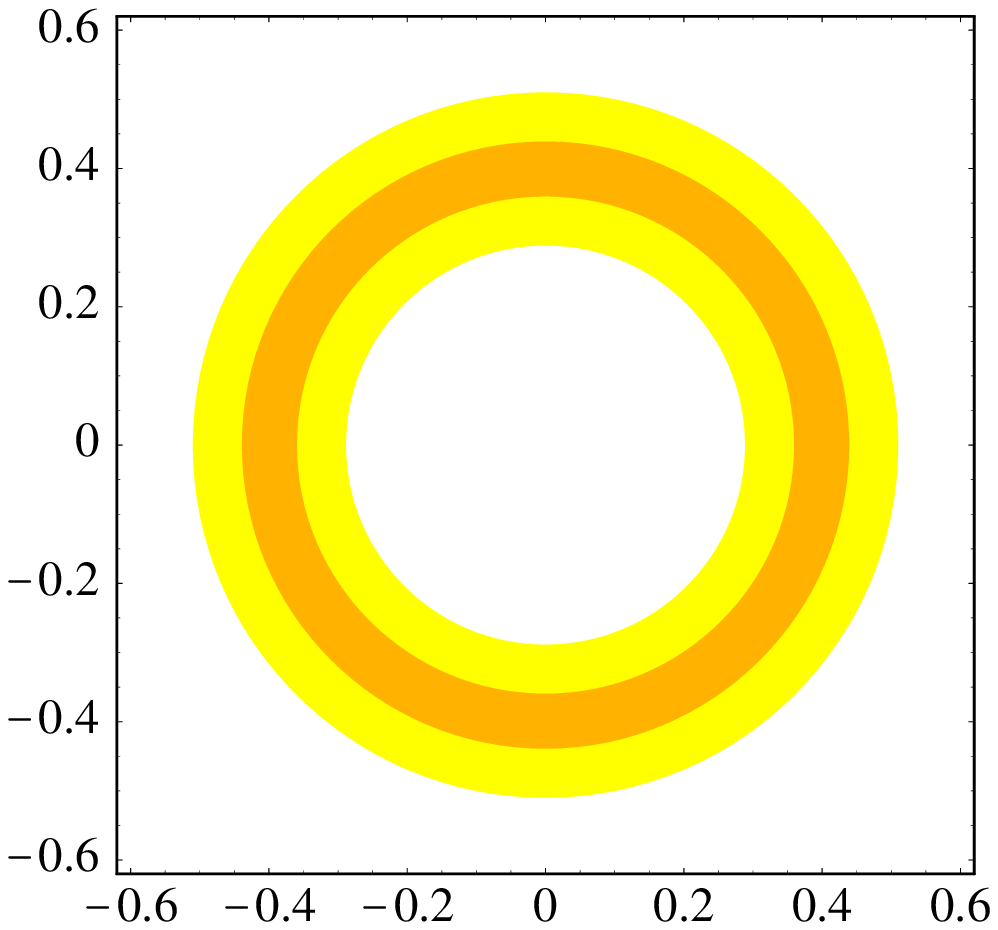}\quad%
\epsfxsize=5cm\epsffile{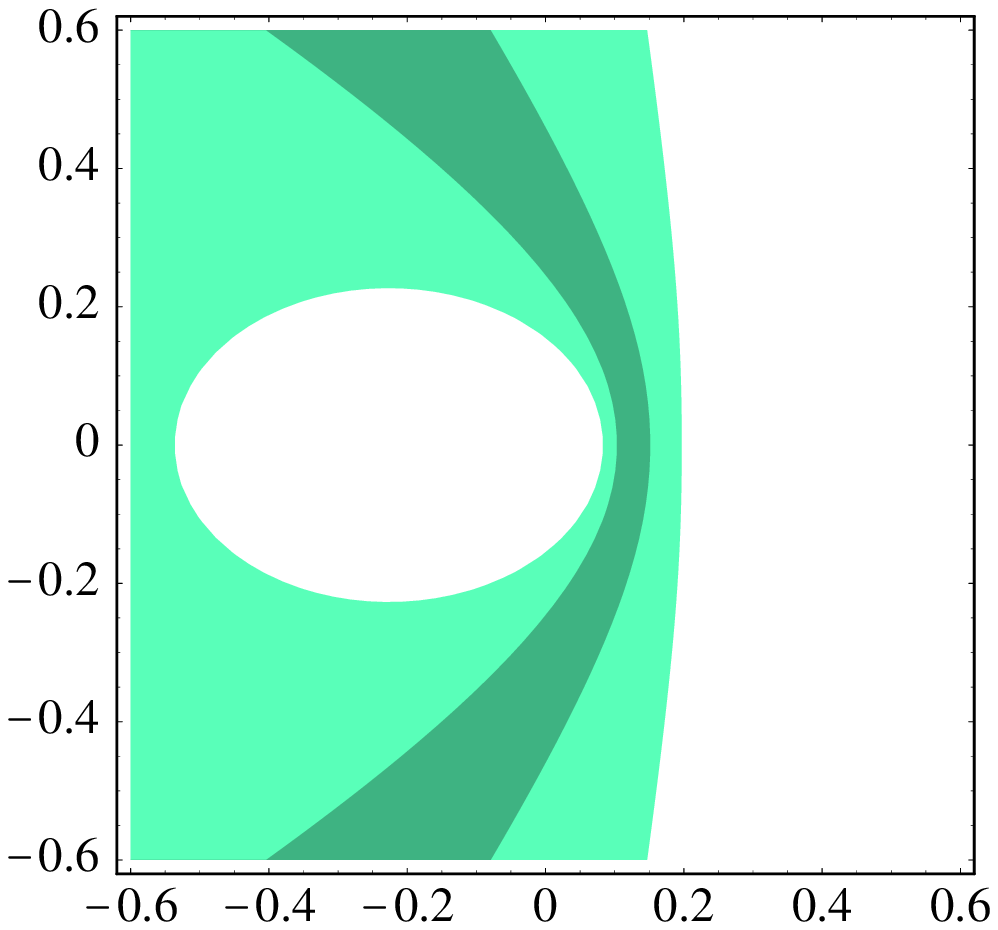}\quad%
\epsfxsize=5cm\epsffile{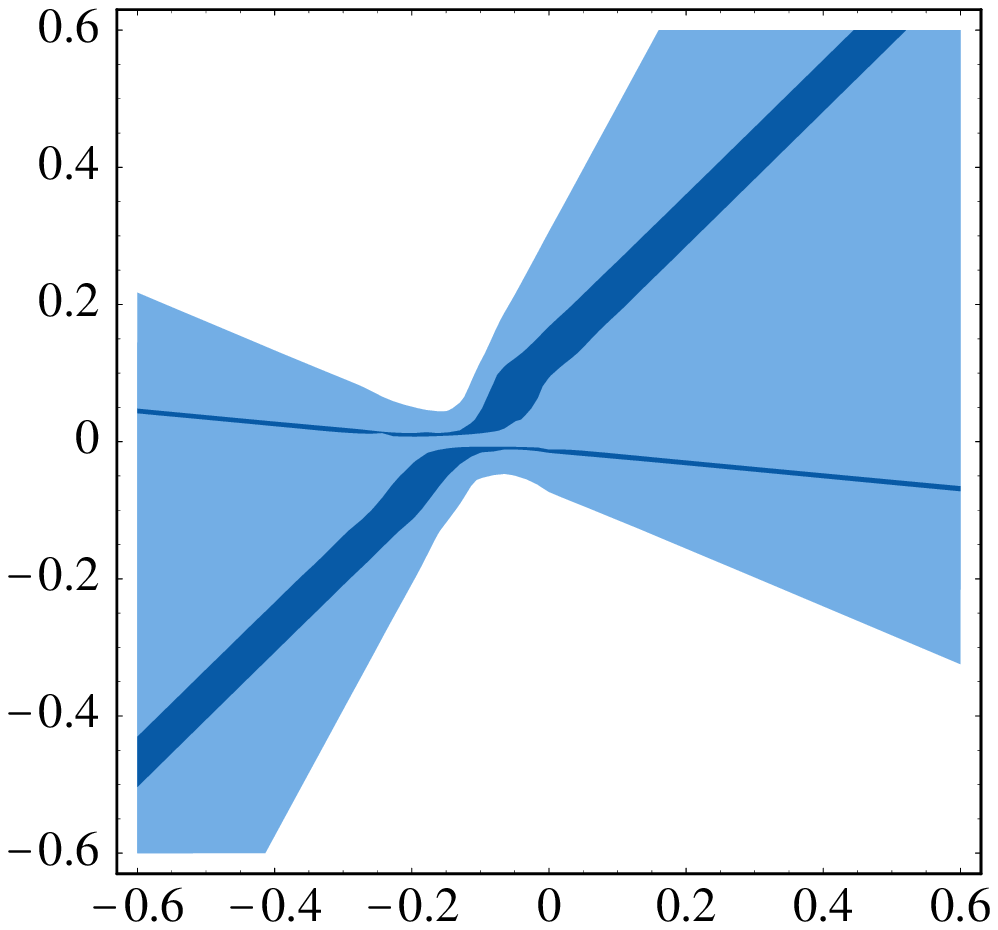}}
\centerline{\parbox{15cm}{\caption{\label{fig:CPT}
The three constraints in the $(\bar\rho,\bar\eta)$ plane used in the 
construction of the CP-$b$ triangle (see text for explanation). 
Experimental errors are shown at $1\sigma$. In each plot, the dark 
band shows the theoretical uncertainty, which is much smaller than the 
experimental error. This shows the great potential of these methods 
once the data will become more precise.}}}
\end{figure}

The first ingredient is the ratio $|V_{ub}/V_{cb}|$ extracted from 
semileptonic $B$ decays, whose current value is 
$|V_{ub}/V_{cb}|=0.090\pm 0.025$. Several strategies have been proposed 
to determine $|V_{ub}|$ with a theoretical accuracy of about 10\%
\cite{Neubert:1993um,Dikeman:1997es,Falk:1997gj,Bauer:2000xf,Neubert:2000ch}, 
which would be a significant improvement. The first plot in 
Figure~\ref{fig:CPT} shows the corresponding constraint in the 
$(\bar\rho,\bar\eta)$ plane. Here and below the narrow, dark-colored 
band shows the theoretical uncertainty, while the lighter band gives the 
current value.

The second ingredient is a constraint derived from the ratio of the
CP-averaged branching fractions for the decays $B^\pm\to\pi^\pm K_S$ and 
$B^\pm\to\pi^0 K^\pm$, using a generalization of the method suggested
in \cite{Neubert:1998pt}. The experimental inputs to this analysis are 
the tree-to-penguin ratio $\varepsilon_{\rm exp}=0.205\pm 0.025$ 
mentioned earlier, and the ratio
\[
   R_* = 
   \frac{\mbox{Br}(B^+\to\pi^+ K^0)+\mbox{Br}(B^-\to\pi^-\bar K^0)}
        {2[\mbox{Br}(B^+\to\pi^0 K^+)+\mbox{Br}(B^-\to\pi^0 K^-)]}
   = 0.78\pm 0.11
\]
of two CP-averaged $B\to\pi K$ branching fractions \cite{newdata}. 
Without any recourse to QCD factorization this method provides a bound 
on $\cos\gamma$, which can be turned into a determination of 
$\cos\gamma$ (for fixed value of $|V_{ub}|/V_{cb}|$) when information 
on the relevant strong-interaction phase $\phi$ is available. I have 
argued at the end of section~\ref{sec:tests} that the phase $\phi$ is 
bound by experimental data (and very general theoretical arguments) to 
be small, of order $(8\pm 10)^\circ$. (In the future, this phase can be 
determined directly from the direct CP asymmetry in 
$B^\pm\to\pi^0 K^\pm$ decays.) It is thus conservative to assume that 
$\cos\phi>0.8$, corresponding to $|\phi|<37^\circ$. With this 
assumption, the corresponding allowed region in the 
$(\bar\rho,\bar\eta)$ plane was analysed in \cite{Beneke:2001ev}, to 
which I refer the reader for details. The resulting constraint is shown 
in the second plot in Figure~\ref{fig:CPT}. 

The third constraint comes from a measurement of the time-dependent
CP asymmetry $S_{\pi\pi}=\sin2\alpha_{\rm eff}$ in $B\to\pi^+\pi^-$ 
decays. The present experimental situation is unfortunately unclear, 
since the measurements by BaBar ($S_{\pi\pi}=-0.01\pm 0.37\pm 0.07$)
and Belle ($S_{\pi\pi}=-1.21_{\,-0.27\,-0.13}^{\,+0.38\,+0.16}$) are
inconsistent with each other \cite{newdata}. The naive average of these 
results gives $S_{\pi\pi}=-0.64\pm 0.42$ (with an inflated error). The 
theoretical expression for the asymmetry is
\[
   S_{\pi\pi} = \frac{2\,\mbox{Im}\,\lambda_{\pi\pi}}
                     {1+|\lambda_{\pi\pi}|^2} \,,
   \quad \mbox{where} \quad
   \lambda_{\pi\pi} = e^{-i\phi_d}\,
    \frac{e^{-i\gamma} + (P/T)_{\pi\pi}}
         {e^{+i\gamma} + (P/T)_{\pi\pi}} \,.
\]
Here $\phi_d$ is the CP-violating phase of the $B_d$--$\bar B_d$ mixing
amplitude, which in the Standard Model equals $2\beta$. Usually, it is
argued that for small $P/T$ ratio the quantity $\lambda_{\pi\pi}$ is
approximately given by $e^{-2i(\beta+\gamma)}=e^{2i\alpha}$, and so 
apart from a ``penguin pollution'' the asymmetry $S_{\pi\pi}\approx
\sin2\alpha$. Here I adopt a different strategy \cite{Beneke:2001ev}. 
In order to become insensitive to possible New Physics contributions to 
the mixing amplitude, I use the experimental value 
$\sin\phi_d=0.78\pm 0.08$ \cite{newdata} and write 
$e^{-i\phi_d}=\pm\sqrt{1-\sin^2\!\phi_d}-i\sin\phi_d$, with a sign 
ambiguity in the real part. (The plus sign is suggested by the standard
fit of the unitarity triangle.) A measurement of $S_{\pi\pi}$ can then 
be translated into a constraint on $\gamma$ (or $\bar\rho$ and 
$\bar\eta$), which remains valid even if the $\sin\phi_d$ measurement 
is affected by New Physics. The result obtained with the current 
experimental values and assuming $\cos\phi_d>0$ is shown in the third 
plot in Figure~\ref{fig:CPT}. The resulting bands for $\cos\phi_d<0$ 
are obtained by a reflection about the $\bar\rho$ axis. This follows
because the expression for $S_{\pi\pi}$ is invariant under the 
simultaneous replacements $e^{-i\phi_d}\to -e^{i\phi_d}$ and 
$\gamma\to -\gamma$.

\begin{figure}
\epsfxsize=10cm
\centerline{\epsffile{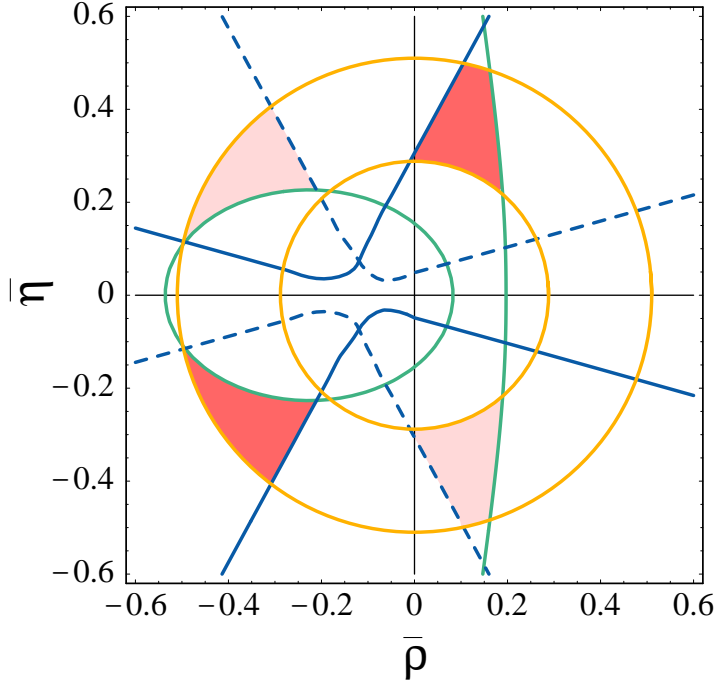}}
\centerline{\parbox{15cm}{\caption{\label{fig:summary}
Allowed regions in the $(\bar\rho,\bar\eta)$ plane obtained from the 
construction of the CP-$b$ triangle. The dashed lines and light-shaded 
areas refer to $\cos\phi_d<0$.}}}
\end{figure}

Each of the three constraints in Figure~\ref{fig:CPT} are, at present, 
limited by rather large experimental errors, while comparison with 
Figure~\ref{fig:UTfit} shows that the theoretical limitations are 
smaller than for the standard analysis. Yet, even at the present level
of accuracy it is interesting to combine the three constraints and 
construct the resulting allowed regions for the apex of the unitarity 
triangle. The result is shown in Figure~\ref{fig:summary}. Note that 
the lines corresponding to the new constraints intersect the circles 
representing the $|V_{ub}|$ constraint at large angles, indicating that 
the three measurements used in the construction of the CP-$b$ triangle 
give highly complementary information on $\bar\rho$ and $\bar\eta$. There 
are four allowed regions, two corresponding to $\cos\phi_d>0$ (dark 
shading) and two to $\cos\phi_d<0$ (light shading). If we use the 
information that the measured value of $\epsilon_K$ requires a positive 
value of $\bar\eta$, then only the two solutions in the upper half-plane 
remain. One of these regions (corresponding to $\cos\phi_d>0$) is close 
to the standard fit, though once again somewhat larger $\gamma$ values 
are preferred. (If only the BaBar result is used for $S_{\pi\pi}$, then 
this region is shifted toward yet larger values of $\gamma$.) I stress 
that this agreement is highly nontrivial, since with the exception of 
$|V_{ub}|$ none of the standard constraints are used in this 
construction. Interestingly, there is a second allowed region 
(corresponding to $\cos\phi_d<0$) which would be consistent with the
constraint from $\epsilon_K$ (see Figure~\ref{fig:UTfit}) and the global 
analysis of charmless hadronic decays (see Figure~\ref{fig:rhoeta}), but 
inconsistent with the constraints derived from $\sin2\beta$ and 
$\Delta m_s/\Delta m_d$. Such a solution would require a significant 
New Physics contribution to $B$--$\bar B$ mixing.

\section{Outlook}

After the by now precise measurement of $\sin2\beta$, the study of 
charmless two-body modes of $B$ mesons is presently the next hottest 
topic in $B$ physics. QCD factorization provides the theoretical 
framework for a systematic analysis of hadronic $B$ decay amplitudes 
based on the heavy-quark expansion. This theory has already passed 
successfully several nontrivial tests, and will be tested more 
thoroughly with more precise data.

A global fit to $B\to\pi K,\pi\pi$ decays establishes the existence of 
a CP-violating phase in the bottom sector of the CKM matrix and tends 
to favor values of $\gamma$ near $90^\circ$, somewhat larger than the 
value suggested by the standard analysis of the unitarity triangle. If 
this trend were real, it would suggest several possibilities for new 
flavor physics beyond the Standard Model, ranging from new contributions 
to $B$--$\bar B$ mixing to non-standard FCNC transitions of the type 
$b\to s g$ or $b\to s\bar q q$. In the future, the construction of the 
CP-$b$ triangle will provide stringent tests of the Standard Model with 
small theoretical uncertainties.

\vspace{0.3cm}  
{\it Acknowledgment:\/} 
This research was supported by the National Science Foundation under 
Grant PHY-0098631. I am grateful to Martin Beneke, Gerhard Buchalla and
Chris Sachrajda for collaboration on much of the work reported here.

\end{document}